\documentclass[aps,prl,superscriptaddress,reprint,showpacs,preprintnumbers,amsmath,amssymb]{revtex4-1}

\usepackage{latexsym}
\usepackage{amsmath}
\usepackage{amssymb}
\usepackage{epsfig}
\usepackage{dcolumn}
\usepackage{bm}
\usepackage{graphicx}
\usepackage{listings}
\usepackage{color}
\usepackage{wasysym}
\usepackage{mathtools}
\usepackage{color}

\begin{document}


\title{Von K{\'a}rm{\'a}n Vortex Street within an Impacting Drop}
\author{Marie-Jean \surname{Thoraval}}
\affiliation{
Division of Physical Sciences and Engineering \& Clean Combustion Research Centre, 
King Abdullah University of Science and Technology (KAUST),
Thuwal, 23955-6900, Saudi Arabia
}

\author{Kohsei \surname{Takehara}}
\author{Takeharu Goji \surname{Etoh}}
\affiliation{
Department of Civil and Environmental Engineering,
Kinki University, Higashi-Osaka, Japan
}

\author{St{\'e}phane \surname{Popinet}}
\affiliation{
National Institute of Water and Atmospheric Research,
Kilbirnie, Wellington 6003, New Zealand
}

\author{Pascal \surname{Ray}}
\author{Christophe \surname{Josserand}}
\author{St{\'e}phane \surname{Zaleski}}
\affiliation{
Institut Jean Le Rond D'Alembert, UMR 7190,
Universit{\'e} Pierre et Marie Curie, Paris, France
}

\author{Sigurdur T. \surname{Thoroddsen}}
\email{sigurdur.thoroddsen@kaust.edu.sa}
\affiliation{
Division of Physical Sciences and Engineering \& Clean Combustion Research Centre, 
King Abdullah University of Science and Technology (KAUST),
Thuwal, 23955-6900, Saudi Arabia
}

\date{\today}

\begin{abstract}
The splashing of a drop impacting onto a liquid pool
produces a range of different sized micro-droplets.
At high impact velocities, the most significant source 
of these droplets is a thin liquid jet emerging at the start 
of the impact from the neck that connects the drop to the pool.
We use ultra-high-speed video imaging in combination 
with high-resolution numerical simulations to show
how the ejecta gives way to \textit{irregular splashing}.
At higher Reynolds number, its base becomes unstable, 
shedding vortex rings into the liquid from the free surface in an axisymmetric 
von K{\'a}rm{\'a}n vortex street, thus breaking the ejecta sheet as it forms.
\end{abstract}

\maketitle

Liquid drop splashing is part of our daily lives,
from the morning shower to natural rain \cite{Rein1993, Yarin2006}.
While it has been studied for more than one hundred years \cite{Worthington1882},
it is only recently that advances in high-speed imaging techniques \cite{Etoh2003, Thoroddsen2008}
have revealed its early dynamics \cite{Thoroddsen2011, Zhang2012}.
Splashing refers herein to the breakup of a drop into smaller droplets during impact.
Understanding the underlying mechanism which produces the smallest droplets
is important for example for the number of microscopic aerosols which remain
when those satellite droplets evaporate. Such aerosols affect human health
and can act as nucleation sites during cloud formation.

For high speed drop impact on a liquid pool,
the \textit{ejecta sheet} is the first stage leading to splashing.
It was first observed in the inviscid numerical simulations of Weiss \& Yarin \cite{WeissYarin1999}
and in the experiments of Thoroddsen \cite{Thoroddsen2002}.
When the drop impacts at higher velocity, the speed of these ejecta sheets increases and they
become thinner. The radial stretching of the sheets reduces their thickness even further
and they can remain intact even at thicknesses well under a micron \cite{Thoroddsen2011}.
When they eventually rupture, they can produce a myriad of very fine spray droplets.
However, this mechanism does not continue for ever,
at a critical Reynolds number the smooth ejecta gives way to a more random splashing,
which counterintuitively may produce fewer small droplets.

\begin{figure}
  \centering
    \includegraphics[width=\linewidth]{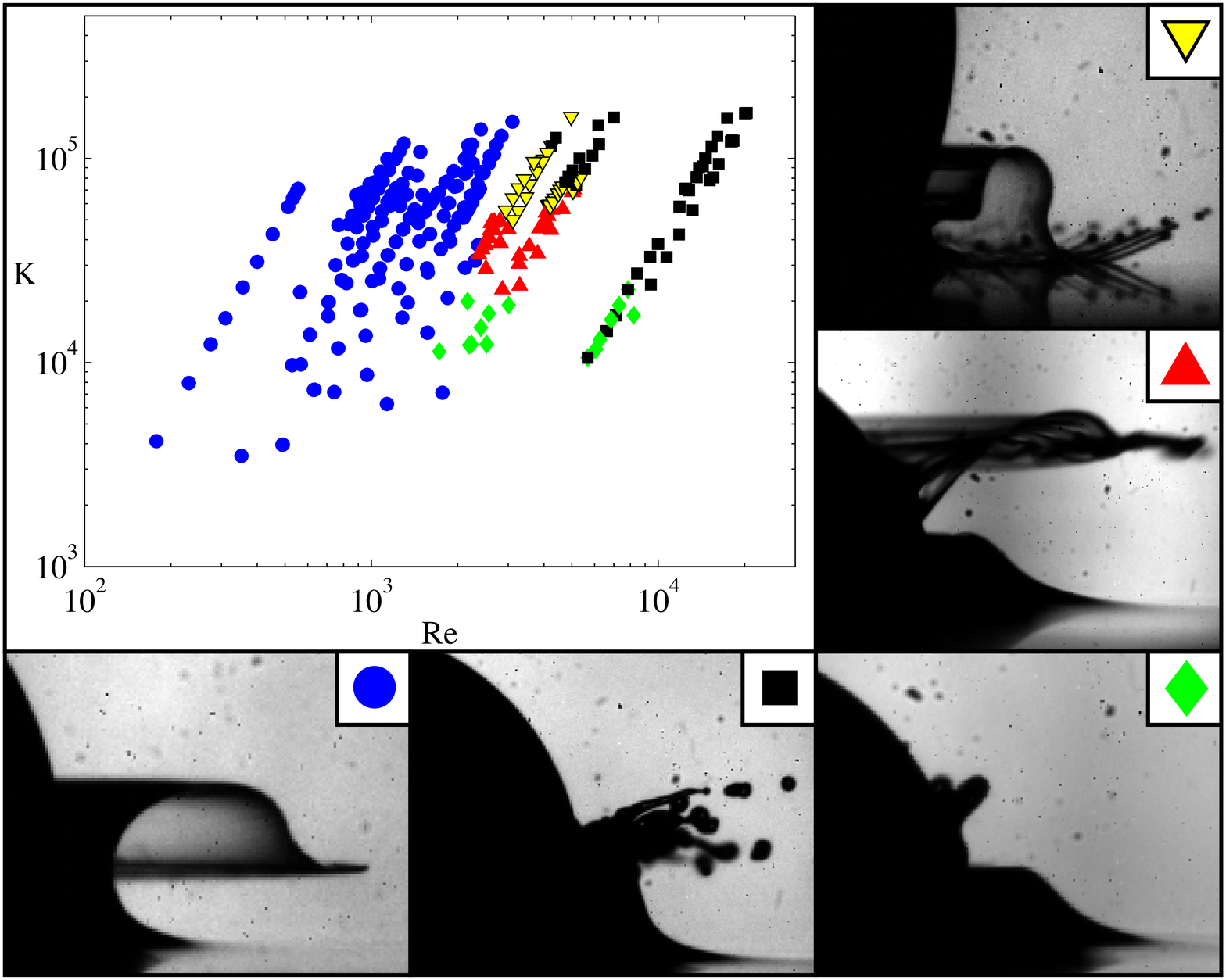}
	\caption{Characterization of the ejecta regimes.
	  (\textcolor{blue}{$\CIRCLE$}) Smooth ejecta sheet.
		($\blacksquare$) Irregular splashing.
		($\mathrlap{\textcolor{yellow}{\blacktriangledown}}\triangledown$) Bumping.
		(\textcolor{red}{$\blacktriangle$}) Quartering.
		(\textcolor{green}{$\blacklozenge$}) Protrusions rising up along the side of the drop.}
  \label{fig:KRe}
\end{figure}

To understand the mechanisms leading from continuous ejecta sheets to \textit{irregular splashing}, 
a systematic study of the early dynamics was conducted with ultra-high-speed video imaging,
over a range of impact velocities $U$, liquid viscosities $\mu$ and droplet diameters $D$ \cite{Supp}.
Figure \ref{fig:KRe} shows a classification of the results in terms of Reynolds number
$Re = \rho D U / \mu$, where $\rho$ is the liquid density,
and splashing parameter $K$, which relates to the Weber number $We = \rho D U^2 / \sigma$, where
$\sigma$ is the surface tension, i.e. $K = We \sqrt{Re}$.
We are interested here in the higher $K$ regime, where splashing occurs \cite{Stow1981, Mundo1995, Cossali1997}.

\begin{figure}
  \centering
	  \includegraphics[width=\linewidth]{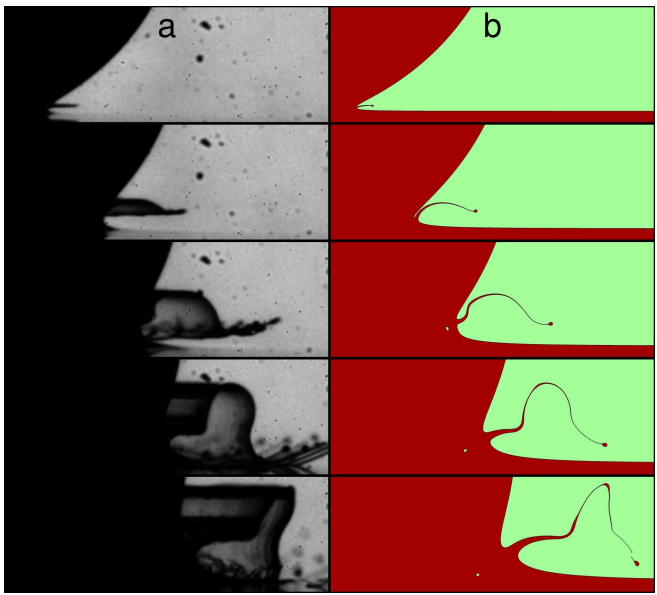}
  \caption{Comparison between experiment and axisymmetric numerical simulation for a bumping case.
	$U = 4.04 m/s, Re = 3.55 \times 10^{3}, K = 7.44 \times 10^{4}$. 
	From top to bottom, observations at time 30, 80, 130, 180 and 230 $\mu s$ after contact. 
	(a) Experimental observation. 
	The static dark points correspond to dust on the camera sensor.
	The video was taken at 200,000 frames per second. 
	(b) Numerical simulation of the drop impact for the same conditions 
	as in the experiment presented in (a).
	In the last image, the leading part of the ejecta sheet becomes smaller
	than the grid size, by stretching between the apex and the tip,
	and thus breaks into non-physical droplets.
	The axisymmetric simulations cannot include the three-dimensional effects,
	such as the breakup of the tip observed in (a).
  Supplemental videos show the two evolutions \cite{Supp}.}
  \label{fig:Bumping}
\end{figure}

The classification in Fig. \ref{fig:KRe} focuses on the ejecta shapes.
In the lower range of $Re$ (more viscous liquids), a smooth ejecta sheet emerges 
between the drop and the pool (\textcolor{blue}{$\CIRCLE$}).
However in the highest range of $Re$, isolated droplets emerge from the neck, 
followed by a disturbed liquid surface, and no coherent ejecta can be identified,
i.e. \textit{irregular splashing} occurs ($\blacksquare$). 
In the intermediate regime ($Re \approx 2000 - 6000 $), 
the ejecta sheets show a large variety of repeatable shapes. 
We have grouped them into 3 classes.
At lower $K$ (lower impact velocities, \textcolor{green}{$\blacklozenge$}), 
surface tension prevents the formation of an ejecta sheet.
However, we observe some protrusions travelling up along the side of the drop,
without ejection of droplets outwards.
At higher $K$ (\textcolor{red}{$\blacktriangle$}), the ejecta sheet is more developed; 
however, it stays attached to the drop, stretching the ejecting sheet between 
the expanding tip of ejecta and the drop entering the pool. 
This stretching can lead to the explosive rupturing of the sheet, 
which generates fast droplets of a large range of sizes through slingshot \cite{Thoroddsen2011}.
In the upper range of $K$ ($\mathrlap{\textcolor{yellow}{\blacktriangledown}}\triangledown$)
we observe an intriguing phenomenon where the free-standing sheet 
interacts strongly with the downwards moving drop surface. 
This is shown in the sequence of Fig. \ref{fig:Bumping}(a),
herein referred to as the \textit{bumping} of the ejecta.
The ejecta is strongly bent by the drop, and then folds at its apex.
Overall snapshot of a bumping ejecta was included in 
Thoroddsen \textit{et al.} \cite{Thoroddsen2008} [their Fig. 8(c)].

Those experimental results clearly show the effect of the Reynolds number
on the transition towards \textit{irregular splashing}.
Moreover, it shows new dynamics of the ejecta sheet interacting with the drop.
It suggests that those interactions could underlie the irregular splashing.

To test this idea, we have chosen to reproduce the impact by numerical simulations.
It is only recently that numerical simulations managed to identify the ejecta sheet
\cite{WeissYarin1999, Josserand2003, Coppola2011}, due to the extreme range of scales involved,
and the challenges of interfacial flow simulations \cite{Tryggvason2011}.
The intricate shapes observed herein were thus not seen numerically until now.
We use the freely available code Gerris \cite{Popinet2009, Agbaglah2011, Gerris},
for its high parallelization and dynamic adaptive grid refinement,
which allow us for the first time to reach enough precision to fully resolve
the dynamics of the ejecta.
This code uses the \textit{volume-of-fluid} method to solve the incompressible Navier-Stokes equations.

Axisymmetric simulations faithfully reproduced all of the experimentally 
observed features, as we demonstrate in Fig. \ref{fig:Bumping} for the bumping case. 
The shape of the drop in the simulation is perfectly spherical,
ruling out the hypothesis that small deviations from spherical drop shapes in the experiments
could be responsible for the drop interaction with the ejecta sheet.

To study this transition to irregular splashing we increase $Re$,
while keeping $K$ constant, from a smooth ejecta sheet
[Fig. \ref{fig:KRe}(\textcolor{blue}{$\CIRCLE$})]
to irregular splashing ($\blacksquare$).
This was done for two different $K$ values, 
corresponding to the bumping ($\mathrlap{\textcolor{yellow}{\blacktriangledown}}\triangledown$) 
and quartering (\textcolor{red}{$\blacktriangle$}) regimes.

\begin{figure}
  \centering
	  \includegraphics[width=\linewidth]{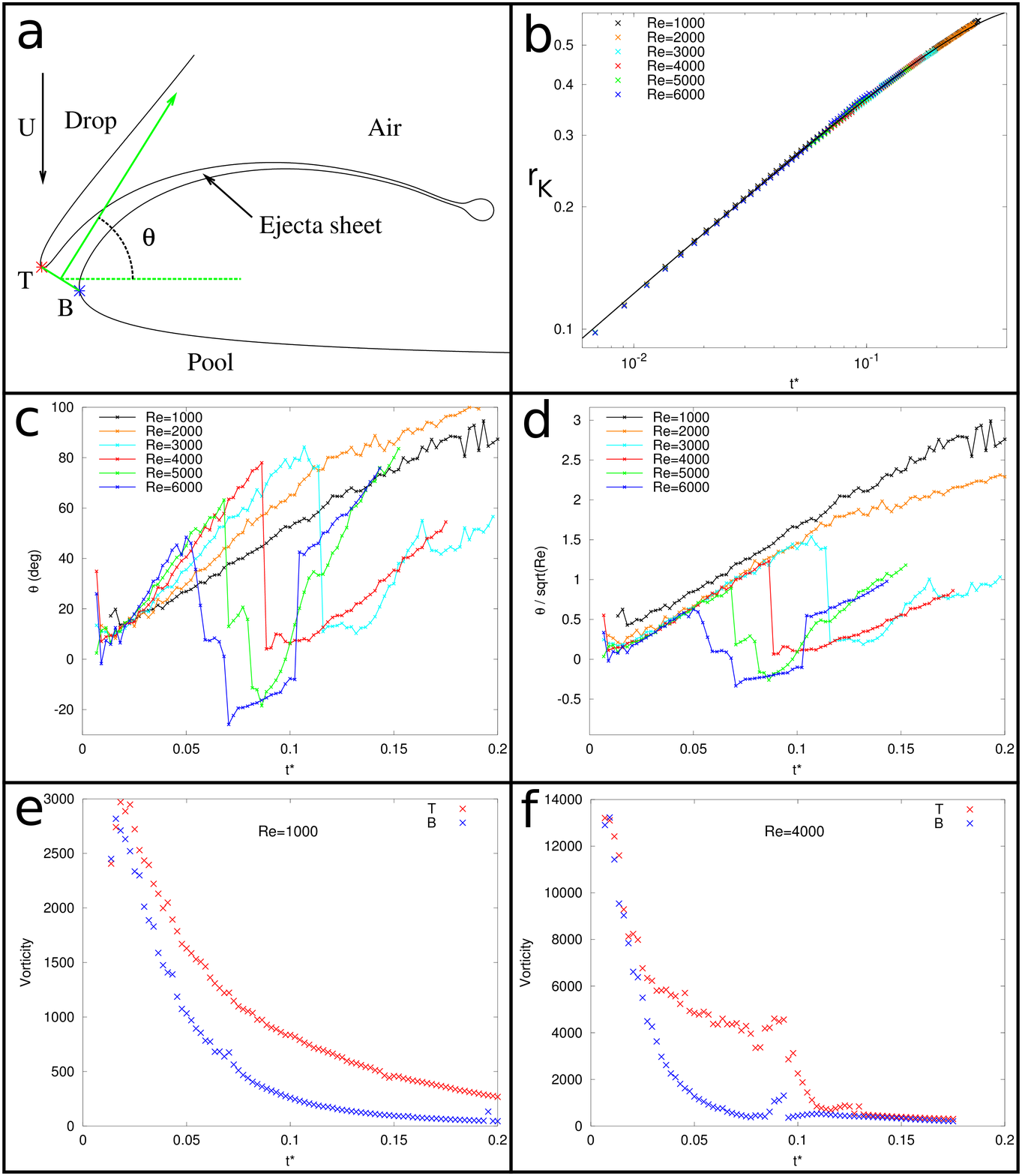}
  \caption{Evolution of the base of the ejecta sheet with $Re$ at $K = 7.44 \times 10^{4}$. 
	Quantities are nondimentionalized by the drop diameter D, 
	the drop impact velocity U and the drop entry time $\tau = D/U$.
	(a) Definition sketch. 
	The base of the ejecta sheet is defined as the segment between the two points of maximum curvature 
	of the interface (T on the drop side and B on the side of the pool). 
	The angle of the ejecta sheet $\theta$ is the angle between the horizontal and the normal to the base. 
	(b) Evolution of the ejecta base radial position, 
	defined as the distance from the axis of symmetry to the middle of TB in (a),
	vs. the nondimensional time $t^{*} = t/\tau$, for $Re$ from 1,000 to 6,000.
	The solid curve is $1.23 r_{J}$, where $r_{J} = \sqrt{t^{*}(1-t^{*})}$ is 
	the radius where an undisturbed drop would meet the pool. 
	(c) Evolution of $\theta$ (in degrees) for $Re$ from 1,000 to 6,000. 
	(d) Same curves as (c), where the angle is scaled by $\sqrt{Re}$. 
	(e,f) Evolution of the maximum positive vorticity (red)
	and maximum absolute negative vorticity (blue) 
	in the liquid near the ejecta base for $Re = 1,000$ (e) and $Re = 4,000$ (f). 
  The positive maximum is located near T, and the negative maximum near B.}
  \label{fig:Base}
\end{figure}

The position of the base of the ejecta $r_{K}$ [Fig. \ref{fig:Base}(a,b)] follows very closely the 
geometric relation predicted by Josserand \& Zaleski \cite{Josserand2003}, 
$r_{K} = C r_{J}$, independent of $Re$, where $r_{J}$ is the radius
where an unperturbed drop would meet the pool surface.
Pure kinematics suggests that the ejecta emerges at an angle intermediate between 
the drop surface and the pool, thus $\theta$ should increase with $r_{J}$.
A simple geometric model \cite{Thoroddsen2011} suggested that $\theta$ increases as 
$\theta \sim \sqrt{t^{*}}$, where $t^{*}$ is the time nondimensionalized by $\tau = D/U$,
whereas the simulations show that $\theta$ grows linearly before bumping [Fig. \ref{fig:Base}(c)].
However, the ejecta rises faster for higher $Re$.
The collapse of the curves in Fig. \ref{fig:Base}(d) shows that 
$\theta$ grows at a rate proportional to $\sqrt{Re}$.
The angle of the ejection velocity vector at the middle of the base also follows a similar trend, 
and increases proportionally to $\sqrt{Re}$.

At lower $Re$, $\theta$ increases slowly enough for the ejecta to escape the drop. 
However, from $Re \gtrsim 3000$, the ejecta sheet rises too fast, thus impacting the drop surface. 
The resulting bumping sharply decreases $\theta$.
This interaction of the drop and the ejecta sheet
observed experimentally occurs earlier at higher Reynolds numbers,
eventually breaking the ejecta sheet.
This is consistent with the interpretation that this interaction
is responsible for the irregular splashing observed at higher $Re$.

Vorticity also appears to play an important role in the dynamics 
of the ejecta sheet [Fig. \ref{fig:Base}(e,f)].
The flow around a curved free surface generates vorticity 
(see for instance Batchelor \cite{Batchelor1967} 5.14).
Therefore, concentrated vorticity is observed in the simulations near 
points T and B at the base of the ejecta, as the flow moves 
around the base to enter the ejecta.
At the early stage of the ejecta formation, both sides of the 
base produce a similar strength of vorticity.
This initial vorticity scales as $\sqrt{Re}$ as observed
previously \cite{Josserand2003}. 
However, the difference in vorticity (absolute values)
between the two sides is observed to increase initially linearly 
with time, before decreasing again.
Moreover, this difference is higher for larger $Re$ [Fig. \ref{fig:Base}(f)].

\begin{figure*}
  \centering
	  \includegraphics[width=0.80\linewidth]{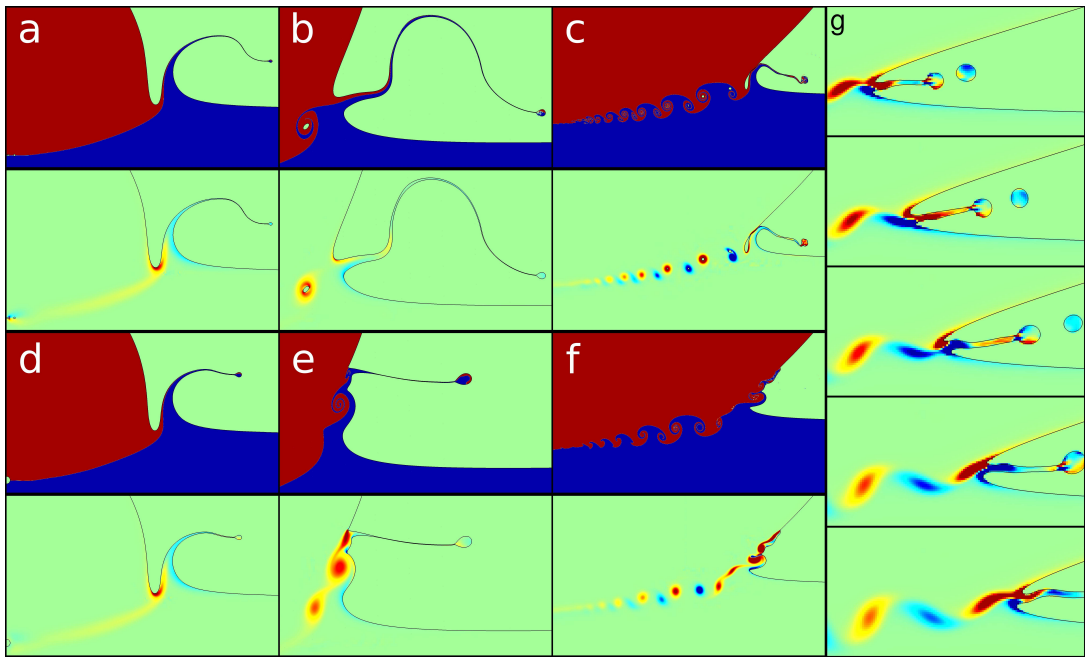}
		\caption{Vorticity structures during drop impact near the transition regime.
		In the top images we differentiate the liquids originating from the drop (red) 
		and from the pool (blue), from the air (green), 
		as can be done in experiments by seeding one or the other with fluorescent dye \cite{Thoroddsen2002}. 
		The bottom images show the corresponding vorticity in the liquid.
		(a-c) Bumping transition, with $K = 7.44 \times 10^{4}$,
		for increasing Reynolds numbers: $Re = 1000$, $3552$ \& $14500$ respectively.
		(d-f) Quartering transition, with $K = 3 \times 10^{4}$,
		for $Re = 1000$, $3552$ \& $10000$ respectively.
		To allow direct comparison, the images of the first row correspond to 
		the same nondimensional time as the ones in the second row,
		with the same field of view.
		(c) and (f) correspond to a water drop impacting at 2.84 m/s
		and 1.98 m/s respectively.
		(g) Details of the early vortex shedding in the same case as (c).
		The period of this shedding shown here is approximately 2 $\mu$s,
		over a radial distance of 50 $\mu$m.}
  \label{fig:vortStructures}
\end{figure*}

By looking closely at the neck region during the impact, 
we can identify fundamental changes in the vorticity structure as $Re$ 
is increased (Fig. \ref{fig:vortStructures}). 
Note that in Fig. \ref{fig:vortStructures}(a) most of the liquid in the sheet 
originates from the pool, in agreement with dye visualizations \cite{Thoroddsen2002}.
For the lower range of $Re$ [Fig. \ref{fig:vortStructures}(a,d)], the vorticity stays concentrated 
near the free surface at the neck of the ejecta sheet.
As there is stronger vorticity generated at the top of the ejecta base,
a vorticity layer of one sign separates the drop and the pool liquids
but it remains stable.
$K$ affects the shape of the outer part of the ejecta sheet,
as we observe by comparing Fig. \ref{fig:vortStructures}(a,d),
consistently with experimental observations \cite{Thoroddsen2011}.
For intermediate $Re$, the interface remains stable in its early evolution.
In the bumping case (b), the rising ejecta sheet contacts the downwards moving drop surface. 
This creates a shear instability, generating a toroidal vortex structure around the entrapped bubble.
In the quartering case (e), the ejecta sheet leaves the neck region to climb up the drop,
pulled by higher surface tension.
This also creates a shear instability between the climbing liquid from the pool and 
the drop liquid moving down, forming a row of vortex rings of the same sign. 
These vortices near the free surface leave their signature \cite{Yu1990} 
by creating waves below the rising sheet, a feature also observed experimentally.
However, all such vortical effects are absent from inviscid theory and simulations
\cite{WeissYarin1999, Howison2005}

At even higher $Re$ [Fig. \ref{fig:vortStructures}(c,f)], vorticity is shed 
behind the base of the ejecta sheet, in a way reminiscent of the von K{\'a}rm{\'a}n 
vortex street, here forming alternating-sign vortex rings.
For the first 7 shedding cycles, the local Reynolds number 
based on the radial speed and width of the neck takes value around 70
and the Strouhal number $St = fD/U$ is around $0.11 \pm 0.05$,
in good agreement with related K{\'a}rm{\'a}n streets.
Moreover, the free surface at the base is also affected by the shedding, 
reinforcing the oscillations \cite{Williamson2004}.
During the early shedding [Fig. \ref{fig:vortStructures}(g)], surface tension effects 
are higher because of the sharper surface geometry.
As the angle of the neck increases, the amplitude of the oscillations increases.
The ejecta can then climb on the drop at lower $K$ (f),
or impact alternatively on the drop and the pool (c),
in a similar way to the bumping, entrapping a row of bubble rings \cite{WeissYarin1999, Davidson2002}.
Four bubble rings can be clearly identified in Fig. \ref{fig:vortStructures}(c),
and a fifth one being created.
Only well resolved bubbles and droplets (larger than 30 cells) are kept in the numerics,
suggesting that smaller bubbles could be entrapped earlier.

%\section*{Conclusions}

From systematic experimental observations, reproduced with axisymmetric simulations, we have detailed 
a new mechanism explaining the irregular splashing of a water drop. 
Previously studied mechanisms have described the droplet separation from the rim of the ejecta 
\cite{Gueyffier1998, WeissYarin1999, Zhang2010}, 
or the destabilization of a liquid sheet \cite{Dombrowski1963, Villermaux2002}. 
Our mechanism however explains the breakup of the ejecta sheet by 
the destabilization of its base, through vortex shedding from the free surface.

\begin{acknowledgements}

The authors are grateful to A. Prosperetti and G. Agbaglah for comments, 
B. Marchand and A. Clo for assistance, T. El Sayed for sharing numerical resources, 
and H. Pottmann for advices on curvature calculations. 
We thank also the Gerris community for their support.
M.-J.T. and S.T.T. were partly supported by KAUST-BERKELEY AEA grant 7000000028.

\end{acknowledgements}

\clearpage
\newpage

\section*{Supplemental Material}
\textbf{Experimental setup:} The liquid viscosity was varied in our experiments
by using glycerin/water mixtures of various mass fractions, 
0, 40, 50, 60, 65, 70, 75 and 80\% of glycerin.
The impact velocity was changed by releasing the drops from between 0.16-2.16 m height.
By varying the circular steel nozzle size, we used four different drop diameters. 
Further details of the experiments can be found in Thoroddsen et al. \cite{Thoroddsen2011}.

\textbf{Numerical simulations:} Direct comparison between the experimental observations and numerical results 
with Gerris \cite{Gerris} were done on a set of different impact conditions.
A convergence study of a bumping case showed that a minimum cell size 1,000 times smaller than the drop diameter 
was necessary to observe the bumping shapes, explaining why previous studies did not observe this evolution. 
However, a much finer grid is needed to reproduce the evolution of very thin ejecta sheets. 
This was accomplished using up to 16 levels of refinement, which means separating the domain 
in two smaller cells 16 times in each direction (the size of the smallest cell is $2^{16}$ times smaller than the domain).
In our simulations, this corresponds to 28,800 cells per drop diameter.
At this level of refinement, an equivalent uniform grid would have more than 4 billion cells ($\left(2^{16}\right)^{2}$),
while our simulation have of the order of 20 million cells.

At the highest level of refinement, the ejecta sheet can become as thin as 500 nm (only 3 cells) 
near the tip, due to the extreme stretching. 
This value is consistent with indirect experimental measurements \cite{Thoroddsen2011}.

We present here results at level 14, on 64 processors,
allowing a systematic investigation of the parameter space.

The curvature of the tracer isolines was estimated by fitting a B{\'e}zier curve of order 3 through 5 points.
For smaller curvature cases, the number of points was increased to 11,
improving the estimate of the position of maximum curvature.

\end{document}